\documentclass[twocolumn,preprintnumbers,amsmath,amssymb,superscriptaddress,aps,prd,longbibliography]{revtex4-2}

\usepackage[pdftex]{graphicx}
\usepackage{xcolor}
\usepackage{amsmath}
\usepackage{pdfpages}
\usepackage{etoolbox} 

\makeatletter
\patchcmd{\@outputpage@head}{\@ifx{\LS@rot\@undefined}{}{\LS@rot}}{}{}{}
\makeatother
\allowdisplaybreaks

\begin{document}


\title{Giant modulation of optical nonlinearity by Floquet engineering}


\author{Jun-Yi Shan}
\affiliation{Department of Physics, California Institute of Technology, Pasadena, California 91125, USA\looseness=-1}
\affiliation{Institute for Quantum Information and Matter, California Institute of Technology, Pasadena, California 91125, USA\looseness=-1}
\author{M. \surname{Ye}}
\affiliation{Kavli Institute for Theoretical Physics, University of California, Santa Barbara, California 93106, USA}
\author{H. Chu}
\affiliation{Department of Physics, California Institute of Technology, Pasadena, California 91125, USA\looseness=-1}
\affiliation{Institute for Quantum Information and Matter, California Institute of Technology, Pasadena, California 91125, USA\looseness=-1}
\author{\textcolor{black}{Sungmin} Lee}
\affiliation{\textcolor{black}{Department of Physics and Astronomy, Seoul National University, Seoul 08826, Republic of Korea\looseness=-1}}
\author{\textcolor{black}{Je-Geun} Park}
\affiliation{\textcolor{black}{Department of Physics and Astronomy, Seoul National University, Seoul 08826, Republic of Korea\looseness=-1}}
\affiliation{\textcolor{black}{Center for Quantum Materials, Seoul National University, Seoul 08826, Republic of Korea\looseness=-1}}
\affiliation{\textcolor{black}{Institute of Applied Physics, Seoul National University, Seoul 08826, Republic of Korea\looseness=-1}}
\author{L. Balents}
\affiliation{Kavli Institute for Theoretical Physics, University of California, Santa Barbara, California 93106, USA}
\author{D. Hsieh}
\email[Corresponding author: ]{dhsieh@caltech.edu}
\affiliation{Department of Physics, California Institute of Technology, Pasadena, California 91125, USA\looseness=-1}\affiliation{Institute for Quantum Information and Matter, California Institute of Technology, Pasadena, California 91125, USA\looseness=-1}

\date{\today}
\maketitle

\textbf{Strong periodic driving with light offers the potential to coherently manipulate the properties of quantum materials on ultrafast timescales. Recently, strategies have emerged to drastically alter electronic and magnetic properties by optically inducing non-trivial band topologies \cite{oka,kitagawa2011,rudner,Hubener2017,mciver,wang}, emergent spin interactions \cite{mentink,claassen,liu,kimel,gil} and even superconductivity \cite{cavalleri}. However, the prospects and methods of coherently engineering optical properties on demand are far less understood \cite{franco}. Here we demonstrate coherent control and giant modulation of optical nonlinearity in a van der Waals layered magnetic insulator, manganese phosphorus trisulfide (MnPS$_3$). By driving far off-resonance from the lowest on-site Mn \textit{d}-\textit{d} transition, we observe a coherent on-off switching of its optical second harmonic generation efficiency on the timescale of 100 femtoseconds with no measurable dissipation. At driving electric fields of order 10$^{9}$ volts per metre, the on-to-off ratio exceeds 10, limited only by the sample damage threshold. Floquet theory calculations \cite{shirley} based on a single-ion model of MnPS$_3$ are able to reproduce the measured driving field amplitude and polarization dependence of the effect. Our approach can be applied to a broad range of insulating materials, raising the possibility of dynamically designed nonlinear optical elements.}

The ability to widely tune the optical nonlinearity of a material with ultrafast speed is crucial for advancing photonics technologies spanning optical signal processing, on-chip nonlinear optical sources and optical computing. However, the nonlinear optical properties of materials, dictated by their electronic and crystallographic structures, are largely set at the synthesis and fabrication stages. Further \textit{in situ} tuning may be achieved by changing temperature, pressure \cite{bayarjargal}, electric field \cite{terhune}, current density \cite{an,ruzicka} or carrier concentration \cite{seyler,soavi}, but these approaches are static and often materials specific. Dynamical tuning based on light-induced phase transitions \cite{satoh,zhang} or photo-carrier density modulation \cite{sartorello} have previously been demonstrated. However, these approaches impart significant heating and are limited in speed due to electronic and structural relaxation bottlenecks.  

Floquet engineering potentially offers a non-thermal and broadly applicable strategy to modulate nonlinearity on ultrashort timescales limited only by the drive pulse duration. However, appreciable tuning requires strong driving (pump) electric fields $E^\mathrm{pu}$ characterized by a Floquet parameter $\mathcal{E}\equiv\frac{eaE^\mathrm{pu}}{\hbar\Omega}$ of order unity, where $e$ is the electron charge, $a$ is the atomic spacing and $\Omega$ is the driving frequency. For a typical solid with $a \sim$ 3 \AA, the requisite field is of order 10$^9$ V/m at optical or near-infrared frequencies, making runaway heating a major obstacle to experimentally realizing Floquet engineering. To mitigate this effect, we focus on driving electrical insulators below their bandgap.    

The layered honeycomb lattice magnetic insulator manganese phosphorus trisulfide (MnPS$_3$) is an ideal demonstration platform for the following reasons. First, it exhibits a large direct bandgap $E_\mathrm{g}$= 3.1 eV in the visible region \cite{piryatinskaya}. Second, the Mn$^{2+}$ moments adopt a Néel antiferromagnetic (AFM) arrangement that breaks the inversion symmetry of its underlying lattice, allowing a finite second-order optical nonlinearity in the electric-dipole (ED) channel. This has recently been detected by optical second harmonic generation (SHG) measurements with an SHG photon energy resonant with $E_\mathrm{g}$ \cite{chu}. Third, the relatively low AFM ordering temperature ($T_\mathrm{N}$ = 78 K) allows thermal versus non-thermal induced effects to be readily distinguished. Fourth, the timescale for spin dynamics, which may be induced by light directly via magneto-optical effects or indirectly via magneto-elastic coupling \cite{magnetoelastic}, is limited to around 5 ps based on the magnetic exchange interaction strength \cite{kurosawa}. Therefore, any dynamics occurring on the timescale of a femtosecond driving pulse can be confined to the charge sector. Lastly, because the Mn 3\textit{d} electrons are highly localized, the optical response and transport properties of MnPS$_3$ are well captured within a single ion picture \cite{grasso}, which enables an analytical derivation of Floquet engineering effects from a microscopic model. 

We first developed a single ion model to understand the AFM order induced static SHG from MnPS$_3$. Owing to the absence of inversion symmetry, this response is dominated by a bulk ED process of the form $P_i(2\omega)=\chi_{ijk}^\mathrm{ED}E_{j}^\mathrm{pr}(\omega)E_{k}^\mathrm{pr}(\omega)$, where the second-order susceptibility tensor $\chi_{ijk}^\mathrm{ED}$ governs the relationship between the incident (probe) electric field $E_{i}^\mathrm{pr}(\omega)$ and the polarization induced at twice the incident probing frequency $P_i(2\omega)$, and the indices $i$, $j$, $k$ run over the \textit{x}, \textit{y}, and \textit{z} coordinates. As shown in the experiments below, we detect exclusively the time-reversal odd (\textit{c}-type) \cite{fiebigreview} component of $\chi_{ijk}^\mathrm{ED}$, which couples linearly to the AFM order parameter. For a near resonant process where $2\hbar\omega\approx E_\mathrm{g}$, the quantum mechanical expression for $\chi_{ijk}^\mathrm{ED(c)}$ is given by (Supplementary section S1) \cite{boyd}  

\begin{equation}\label{eqn:MnPS3_1}
\begin{split}
    \chi_{ijk}^\mathrm{ED(c)}\propto\sum&\frac{\langle i\vert r_i\vert f\rangle\langle f\vert r_j\vert m\rangle\langle m\vert r_k\vert i\rangle}{(E_f-E_i-2\hbar\omega-i\gamma_f)(E_m-E_i-\hbar\omega)}\\&+(j\leftrightarrow k)
    \end{split}
\end{equation}

\noindent where the sum is performed over Mn$^{2+}$ ions in a unit cell, $\vert i\rangle$, $\vert m\rangle$, and $\vert f\rangle$ are the ground, intermediate, and final states of the SHG process, $E_i$, $E_m$ and $E_f$ denote their respective energies and $\gamma_f$ is a phenomenological decay rate of the final state (Supplementary section S2). In the presence of an octahedral crystal field imposed by the sulfur ions, the five-fold degenerate Mn 3\textit{d} orbitals split into a low energy $t_{2g}$ triplet and a high energy $e_g$ doublet. The ground state is a high-spin (\textit{S} = 5/2) state characterized by a $t_{2g}^3e_g^2$  orbital configuration with $ ^6A_{1g}$ symmetry. According to prior optical absorption measurements (Fig. 1a) \cite{grasso}, the intermediate state has predominantly $^4T_{1g}(t_{2g}^4 e_g^1)$ character (\textit{S} = 3/2) and the final state has predominantly S 3\textit{p} $\rightarrow$ Mn 3\textit{d} charge transfer (CT) character (\textit{S} = 5/2). The $\vert f\rangle$ state has opposite parity to the $\vert i\rangle$ and $\vert m\rangle$ states. By introducing spin-orbit coupling $\lambda$ and a trigonal distortion of the crystal field $\eta$ as perturbations to the states described above \cite{muthukumar}, optical transitions $\vert i\rangle\rightarrow\vert m\rangle$ and $\vert m\rangle\rightarrow\vert f\rangle$ become ED allowed (Fig. 1b). Upon coherently summing the single ion contributions from two Mn$^{2+}$ sites in the unit cell, one obtains $\chi_{ijk}^\mathrm{ED(c)}\propto\beta_{ijk}\lambda^2\eta(\langle S_{z,1}\rangle-\langle S_{z,2}\rangle)$, where $(\langle S_{z,1}\rangle-\langle S_{z,2}\rangle)$ is the staggered moment perpendicular to the honeycomb plane. The coefficient $\beta_{ijk}$ encodes the symmetry of the underlying crystal through the matrix elements in Eqn. 1. To capture the loss of three-fold rotational symmetry due to coupling between adjacent honeycomb layers displaced along \textit{x}, we assign unequal weight to the dipole matrix elements along \textit{x} and \textit{y}.

To verify this static SHG model, we performed rotational anisotropy (RA) measurements \cite{harter} using near resonant probe light ($\hbar\omega$ = 1.55 eV). The beam was focused obliquely onto a bulk MnPS$_3$ single crystal and specular reflected SHG light was collected as a function of the scattering plane angle $\varphi$ (Fig. 1c). Above $T_\mathrm{N}$ we observe a weak temperature independent SHG signal arising from \textit{i}-type (time-reversal even) higher multipole bulk crystallographic SHG processes (Fig. 1d), consistent with a previous report \cite{chu}. Below $T_\mathrm{N}$ the intensity, collected at $\varphi=60^{\circ}$, undergoes a steep upturn that can be fitted to a power law $\chi_{ijk}^\mathrm{ED(c)}\propto(T_\mathrm{N}-T)^{\beta}$ with $\beta$ = 0.32. This is in excellent agreement with the critical exponent of the AFM order parameter ($\beta$ = 0.32) obtained from neutron diffraction \cite{wildes} (Supplementary section S3), confirming its linear coupling to $\chi_{ijk}^\mathrm{ED(c)}$ as predicted in our model. The enhanced anisotropy of the RA pattern below $T_\mathrm{N}$ arises from interlayer coupling and is fully captured in our model through the $\beta_{ijk}$ coefficient (Fig. 1d inset). 

Next, we study how an electric field oscillating at sub-gap frequency affects the electronic spectrum of MnPS$_3$ within our single ion model. Since this drive mainly hybridizes $\vert i\rangle$ and $\vert f\rangle$ due to their opposite parity and equal spin, the three-level problem can be simplified to a two-level one, described by the following time-dependent Hamiltonian   
\begin{equation}
    H(t)=H_0+e\vec{r}\cdot\vec{E}^\mathrm{pu}\cos\Omega t
\end{equation}

\noindent where $H_0$ is the un-perturbed 2 $\times$ 2 Hamiltonian. By diagonalizing the time-independent Floquet Hamiltonian \cite{rudner} $(H_f)_{mn}=\frac{1}{2\pi/\Omega}\int_0^{2\pi/\Omega}e^{i\Omega t(m-n)}H(t)-n\hbar\Omega\delta_{mn}$ truncated at the $\pm3^{\textrm{rd}}$ Floquet sector (Fig. 2a inset), we obtain the pump field dressed initial and final states $\vert i'\rangle$ and $\vert f'\rangle$    

\begin{equation}
\begin{split}
    \vert i'\rangle&=e^{-i(E_i-\Delta E)t/\hbar}(\cos\alpha\,\vert i\rangle+\sin\alpha\,e^{i\phi(t)} \vert f\rangle)\\
    \vert f'\rangle&=e^{-i(E_f+\Delta E)t/\hbar}(-\sin\alpha\,e^{-i\phi(t)} \vert i\rangle+\cos\alpha\,\vert f\rangle)
\end{split}
\end{equation}

\noindent where $\Delta E$ is the energy shift and the hybridization is parameterized by a mixing amplitude $\sin\alpha$ and phase $\phi(t)$, which all depend on $\vec{E}^\mathrm{pu}$ (Supplementary section S4). For a Gaussian pulsed drive, our calculations show that in the adiabatic limit where the pulse width far exceeds $\Omega^{-1}$, both the bandgap and hybridization undergo a temporal increase that follows the pulse envelope (Fig. 2a), attaining maximum values at the peak pump field $E_\mathrm{max}^\mathrm{pu}$. The maximal mixing amplitude scales linearly with $E_\mathrm{max}^\mathrm{pu}$ as expected from a perturbative treatment, whereas the maximal bandgap increase ($2\Delta E$) scales like the square of $E_\mathrm{max}^\mathrm{pu}$. While this quadratic dependence is reminiscent of the optical Stark effect \cite{autler,siestark}, our Floquet treatment goes beyond the rotating wave approximation by including both optical Stark and Bloch-Siegert shifts \cite{bs} (Fig. 2b) as well as the influence of higher Floquet sectors, predicting $2\Delta E_\mathrm{max}$ as large as 188 meV for $E_\mathrm{max}^\mathrm{pu}=10^9$ V/m.  

Both mixing and bandgap widening, imparted by a coherent modulation of the two-level Hamiltonian comprised of $\vert i\rangle$ and $\vert f\rangle$, should suppress the magnitude of $\chi_{ijk}^\mathrm{ED(c)}$ because the former reduces the amplitude of states in the zeroth Floquet sector – the dominant contribution to $\chi_{ijk}^\mathrm{ED(c)}$ – by a factor of $\cos\alpha$, while the latter shifts the resonance condition away from $\hbar\omega$ = 1.55 eV. The fast oscillating pump field induces a quasi-static change in the time-averaged value of $\chi_{ijk}^\mathrm{ED(c)}$ that follows the slower pump pulse envelope, consistent with a Floquet description. To quantify these effects, we computed the expected change in $\chi_{ijk}^\mathrm{ED(c)}$ and the resulting modulation of the magnetic contribution to the SHG intensity $I^\mathrm{mag}$ (Fig. 1d) within our single ion model using the dressed initial and final states, assuming $\hbar\Omega$ well below the $^6A_{1g}$ $\rightarrow$ $^4T_{1g}$ transition and $\vec{E}^\mathrm{pu}$ parallel to the nearest neighbor Mn-Mn bond ($\theta=90^{\circ}$). As shown in the inset of Figure 2b, we predict an inverse power law-like dependence of $I^\mathrm{mag}$ on the driving field amplitude, indicating that the suppression is predominantly caused by energy shifts that affect the denominator in Eqn. 1. Remarkably, we predict that Floquet engineering can impart a giant suppression exceeding 90 \% at readily attainable field strengths of order 10$^9$ V/m.   

To experimentally test our prediction, we performed time-resolved pump-probe RA-SHG measurements in the AFM phase of MnPS$_3$. To minimize dissipation and decoherence, the pump photon energy was tuned below the $^6A_{1g}$ $\rightarrow$ $^4T_{1g}$ transition edge near 2 eV to avoid absorption, but above 0.5 eV to suppress the effects of quantum tunneling between valence and conduction bands, phonon resonances, and photo-assisted inter-site hopping (Supplementary section S8) that are more pronounced at lower frequencies. Gaussian pump and probe pulse envelopes of 120 fs and 80 fs duration were used respectively, satisfying the adiabatic condition. Figure 3a shows instantaneous RA patterns at select time delays measured using $\theta=90^{\circ}$ and $E_\mathrm{max}^\mathrm{pu}=10^9$ V/m. The magnitude of the RA patterns is drastically reduced during pumping and can be fit by simply decreasing all $\chi_{ijk}^\mathrm{ED(c)}$ elements uniformly. The temporal evolution of the RA patterns is completely symmetric about time \textit{t} = 0 – the instant when pump and probe pulses are exactly overlapped – and the transient SHG intensity change $\Delta I^\mathrm{mag}/I^\mathrm{mag}$ exhibits a temporal profile that matches the theoretically predicted SHG profile convolved with the probe pulse (Fig. 3b). These data indicate a coherent and uniform modulation of the $2'/m$ magnetic point group allowed $\chi_{ijk}^\mathrm{ED(c)}$ elements with no measurable dissipation (Supplementary section S6), in accordance with a Floquet engineering process. The maximal suppression of $I^\mathrm{mag}$ reaches around 90 \% and is unchanged upon sweeping $\hbar\Omega$ from 0.66 eV to 1.55 eV, in full agreement with our theoretical model (Supplementary section S4). 

In contrast, measurements performed with $\hbar\Omega$ tuned near the $^6A_{1g}$ $\rightarrow$ $^4T_{1g}$ absorption peak reveal dynamics that are strongly asymmetric about \textit{t} = 0. Following an initial fast coherent reduction of $I^\mathrm{mag}$, there is a slow exponential decay to 100 \% suppression, where it remains for more than 500 ps (Fig. 3d). The decay and plateau are consistent with an incoherent quasi-thermal melting of the AFM order via heat transfer from the optically excited electronic subsystem to the spin subsystem, followed by a very slow cooling of the pumped region through diffusion (Supplementary section S7). This interpretation is further corroborated by instantaneous RA data acquired within the exponential decay time window, which directly map onto our temperature dependent RA data (Fig. 3c).

To directly confirm the predicted bandgap widening effect (Fig. 2), we performed transient SHG spectroscopy measurements with $\hbar\Omega$ = 0.66 eV. The equilibrium SHG spectrum exhibits a steep intensity upturn near the band edge of MnPS$_3$ at $2\hbar\omega$ = 3.05 eV (Fig. 3e), in accordance with the optical absorption spectrum \cite{piryatinskaya} as expected. Upon driving, the band edge feature instantaneously shifts to higher energy, which is opposite to the typical response of electronic gaps to photo-excitation. The size of the positive shift at $t$ = 0 increases monotonically with $E_\mathrm{max}^\mathrm{pu}$ and agrees reasonably well with our theoretically predicted values (Fig. 3f \& g), further supporting the Floquet engineering interpretation.

As both the bandgap widening and level mixing are dependent on the Rabi frequency $\langle f\vert e\vec{r}\cdot\vec{E}^\mathrm{pu}/\hbar\vert i\rangle$, we expect the magnitude of SHG modulation to be tunable by both the electric field amplitude and polarization of the pump pulse. To study this relationship, we performed a comprehensive experimental mapping of $\Delta I^\mathrm{mag}/I^\mathrm{mag}(t=0)$ as a function of both $E_\mathrm{max}^\mathrm{pu}$ and $\theta$ using $\hbar\Omega$ = 0.66 eV (Fig. 4a). A comparison to our model calculation performed over the same parameter space (Fig. 4b), using the same weighting of dipole matrix elements along \textit{x} and \textit{y} as in our static model to account for inter-layer coupling, shows excellent agreement in overall trend. More detailed comparisons can be drawn by taking different one-dimensional cuts through our data set. For a fixed $\theta$, $\Delta I^\mathrm{mag}/I^\mathrm{mag}$ exhibits an expected inverse power law-like dependence on pump field in both experiment and theory (Fig. 4c), with good agreement on the level of suppression.  For a fixed pump field, we observe a sinusoidal dependence of $\Delta I^\mathrm{mag}/I^\mathrm{mag}$ on $\theta$ that is reproduced in our calculations (Fig. 4d). Although the three-fold rotational symmetry of an isolated honeycomb layer forbids an anisotropic Rabi frequency, this is broken in bulk MnPS$_3$ due to the layer stacking (inset Fig. 1d), resulting in a maximum (minimum) Rabi frequency at $\theta=90^{\circ}$ (0$^\circ$). The fact that the $\theta$ dependence remains largely unchanged upon rotating $\phi$ (Fig. 4e) confirms that the anisotropy is intrinsic to the crystal and is unrelated to the relative polarization of the pump and probe light. The close agreement between our measurements and theoretical calculations, which contain no free parameters, confirms the validity of our single-ion treatment and highlights its dominant role over photo-assisted inter-site hopping effects in our experiments (Supplementary section S8).

The Floquet engineering strategy demonstrated here can be broadly applied to coherently control a variety of nonlinear optical processes including optical rectification and higher harmonic generation. Moreover, both coherent enhancement and suppression of the nonlinear response can in principle be realized by tuning the probe photon energy to either side of an absorption resonance. Introducing few-layer exfoliable materials like MnPS$_3$ into cavity architectures \cite{sentef} raises the further exciting prospect of coherently switchable optical, optoelectronic and magnetic devices with reduced external field thresholds.

\section*{M\lowercase{ethods}}
\subsubsection*{Sample preparation}
Single crystals of MnPS$_3$ were synthesized by a chemical vapor transport method using the starting materials; manganese powder (99.95\%, Alfa Aesar), red phosphorous (99.99\%, Sigma-Aldrich), and sulfur flakes (99.99\%, Sigma-Aldrich) were mixed in the stoichiometric ratio with 5 wt\% of extra sulfur within an Ar atmosphere ($<$1 ppm of moisture and oxygen). We put a quartz ampoule containing the raw materials into a horizontal 2-zone furnace with a temperature difference of 780 $^{\circ}$C (hot zone) and 730 $^{\circ}$C (cold zone) for 7 days. The quartz tube was cooled to room temperature over 2 days. We annealed single crystals for an additional one day under Ar atmosphere to remove extra sulfur. We verified the sample stoichiometry with energy dispersive x-ray spectroscopy and carried out the magnetization measurement using a SQUID magnetometer (Quantum Design, MPMS3) (Supplementary section S9). Prior to optical measurements, crystals were cleaved along the (001) planes and then immediately pumped down to a pressure better than 10$^{-7}$ Torr. The results were reproduced in two different samples from two growth batches.

\subsubsection*{Time-resolved SHG measurements}
We used a Ti:sapphire laser with a repetition rate of 1 kHz. The fundamental output of the laser at 800 nm was used as the probe pulse (80 fs width), which was focused obliquely onto a 60 $\mu$m spot on the cleaved surface of the MnPS$_3$ crystal at a 10$^{\circ}$ angle of incidence with a fluence of 1.4 mJ/cm$^2$. The RA patterns were obtained using a fast-rotating scattering plane based technique \cite{harter}. Part of the fundamental output was split off to an optical parametric amplifier to generate the pump pulse (duration 120 fs and bandwidth 60 nm). The pump pulse was focused normally onto an 80 $\mu$m spot on the sample. The peak pump electric field was kept below $1.25\times10^9$ V/m (25 mJ/cm$^2$) to avoid sample damage. The time-resolved measurements were carried out at 10 K unless otherwise stated. For each data point we took the average of four independent measurements to reduce read-out noise from the CCD detector. The integration times used were 900 s per RA pattern and 30 s per data point for single-angle measurements. All reported sizes, widths, and durations are full width at half maximum (FWHM).

\subsubsection*{Transient SHG spectroscopy measurements}
Part of the fundamental output of the laser at 800 nm was directed into a second optical parametric amplifier for the SHG spectroscopy measurements. The output from this optical parametric amplifier was frequency doubled using a BBO crystal in order to generate incident probe beams from 1.48 - 1.63 eV. The power of the incident beam on the sample was kept constant for different photon energies in this range. Wavelength-dependent variations in the quantum efficiency of the CCD detector and in the transmission through the spectral filters were corrected for in the presented data. For each probe energy, the EQ contribution ($\approx 24 \%$ of the total signal) was subtracted off to isolate the $I^\mathrm{mag}$ contribution.

\section*{A\lowercase{cknowledgements}}

We acknowledge discussions with X. Li, S. Chaudhary,  and G. Refael. This work was supported by ARO MURI Grant No. W911NF-16-1-0361. D.H. also acknowledges support for instrumentation from the David and Lucile Packard Foundation and from the Institute for Quantum Information and Matter, an NSF Physics Frontiers Center (PHY-1733907). M.Y. acknowledges support by the Gordon and Betty Moore Foundation through Grant GBMF8690 to UCSB and by the National Science Foundation under Grant No.\ NSF PHY-1748958. J.G.P. was supported by the Leading Researcher Program of the National Research Foundation of Korea (Grant No. 2020R1A3B2079375). 

\section*{C\lowercase{ontributions}}
S.L. and J.G.P. synthesized and characterized the MnPS$_3$ crystals. J.S. and H.C. performed the optical measurements. M.Y., J.S. and L.B. performed the single-ion model based static and Floquet dynamical calculations. J.S., M.Y. and D.H. wrote the paper with input from all authors.

\bibliography{citations}

\begin{figure*}[hbt!]
\includegraphics[width=2\columnwidth]{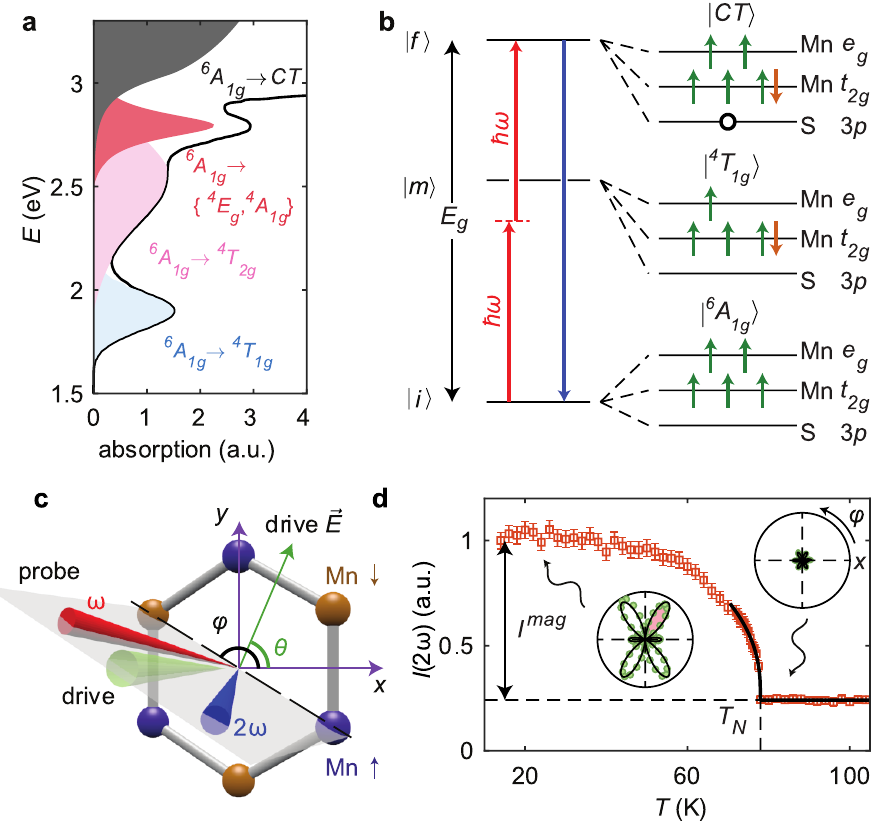}
\caption{\label{fig1} \textbf{Static SHG from MnPS$_3$. a,} Absorption spectrum of MnPS$_3$ (black curve) and underlying optical transitions (shaded areas) adapted from Ref. \cite{grasso}. \textbf{b,} Left: Black horizontal lines denote the initial, intermediate, and final multi-electron states involved in the resonant ED-SHG process (red and blue arrows). Right: Orbital and spin configurations of the states. \textbf{c,} Depiction of the experimental geometry and the antiferromagnetic spin arrangement on the Mn sublattice. The probe beam (red) is focused obliquely onto the sample, and the reflected SHG beam (blue) is measured as a function of the scattering plane angle $\varphi$. Both incident and reflected beams are linearly polarized in the scattering plane. The driving beam (green) is focused normally onto the sample with linear polarization along $\theta$. The \textit{x} and \textit{y} axes correspond to the crystallographic \textit{a} and \textit{b} axes. \textbf{d,} Temperature-dependent SHG intensity acquired at $\varphi=60^{\circ}$ (pink lobe in inset) normalized by its value at 10 K. The error bars represent the standard errors of the mean from four independent measurements. A power law fitting with $\beta$ = 0.32(2) is overlaid (black curve). Insets show RA-SHG patterns above and below $T_\mathrm{N}$ (green circles) and fits to the single ion model (black curves).}
\end{figure*}

\begin{figure*}[hbt!]
\includegraphics[width=2\columnwidth]{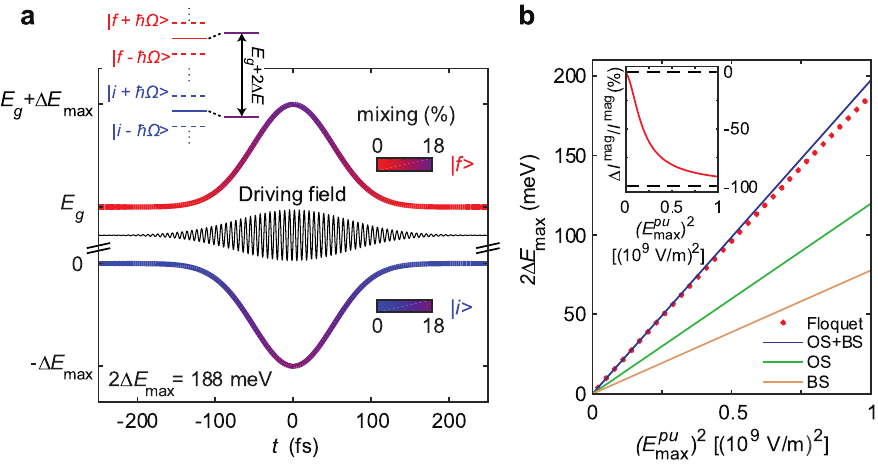}
\caption{\label{fig2} \textbf{Coherent drive-induced state modification. a,} Calculated temporal profile of the energy shift and mixing amplitude of the initial (blue) and final (red) states of the ED-SHG process due to a pulsed periodic drive (black curve). We assumed a Gaussian envelope of width 120 fs, a peak driving field $E_\mathrm{max}^\mathrm{pu}=10^9$ V/m, and polarization $\theta=90^{\circ}$. Inset shows the higher Floquet sectors (dashed lines) that hybridize with the states in the zeroth Floquet sector (solid lines). \textbf{b,} Predicted maximum energy shift versus peak driving field calculated using our full Floquet formalism (Floquet), optical Stark shift (OS), Bloch-Siegert shift (BS), and the sum of OS and BS. Inset shows a calculation of the corresponding change in the magnetic contribution to the SHG intensity (defined in Fig. 1d).}
\end{figure*}

\begin{figure*}[hbt!]
\includegraphics[width=2\columnwidth]{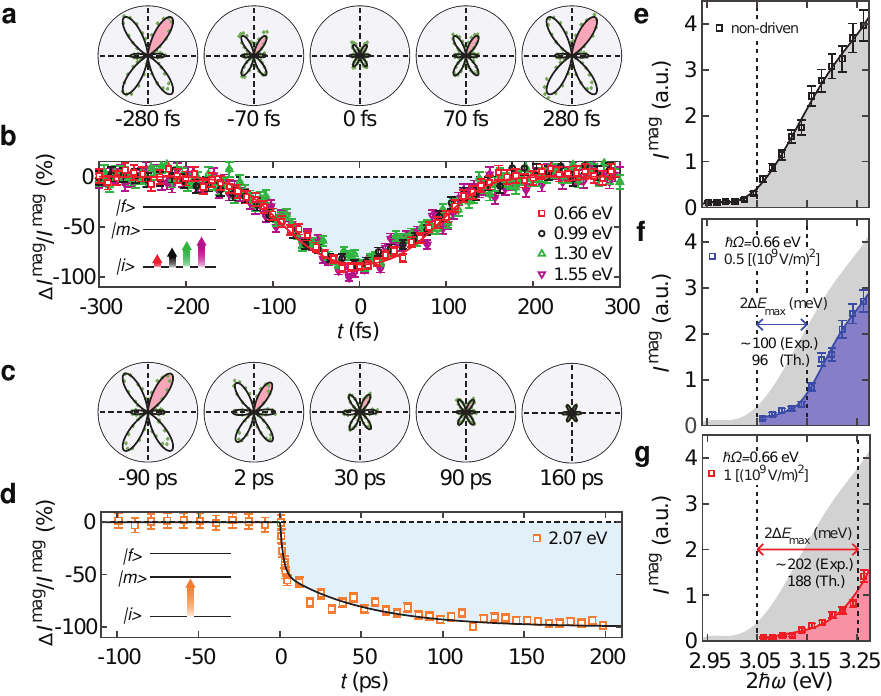}
\caption{\label{fig3} \textbf{Driving photon energy dependence of RA-SHG transients. a,} Time-resolved RA-SHG patterns from MnPS$_3$ measured at 10 K following a pulsed sub-gap drive with $\hbar\Omega$ = 0.66 eV and $E_\mathrm{max}^\mathrm{pu} = 10^9$ V/m (green circles). Black curves are fits to our Floquet model. \textbf{b,} $\Delta I^\mathrm{mag}/I^\mathrm{mag}$ transients measured at $\varphi=60^{\circ}$ (pink lobes in panel \textbf{a}) for different sub-gap pump photon energies (inset) and $E_\mathrm{max}^\mathrm{pu}$ fixed at 10$^9$ V/m. The red curve shows the theoretically predicted SHG response for 0.66 eV drive convolved with the probe pulse profile. Pump-induced changes in the linear optical response or competing second-order nonlinear processes can be excluded as the cause of SHG suppression (Supplementary section S5). \textbf{c,} Time-resolved RA-SHG patterns measured under resonant pumping ($\hbar\Omega$ = 2.07 eV) conditions (inset) with $E_\mathrm{max}^\mathrm{pu}$ set to $7.5\times10^8$ V/m (green circles). Fits to the static RA patterns (Fig. 1d) at temperatures 10 K, 64 K, 76 K, 77 K and 80 K (left to right) are overlaid for comparison. \textbf{d,} Corresponding $\Delta I^\mathrm{mag}/I^\mathrm{mag}$ transient for resonant pumping conditions. Black curve is a guide to the eye. \textbf{e,} Non-driven SHG spectrum at 10 K. \textbf{f and g,} Transient $t$ = 0 SHG spectra at two different $E_\mathrm{max}^\mathrm{pu}$ values. Solid curves in panels e)-g) are guides to the eye. Vertical dashed lines mark the intensity upturns. Values of the observed and theoretically calculated energy shifts are indicated. All error bars represent the standard errors of the mean from four independent measurements.}
\end{figure*}

\begin{figure*}[hbt!]
\includegraphics[width=2\columnwidth]{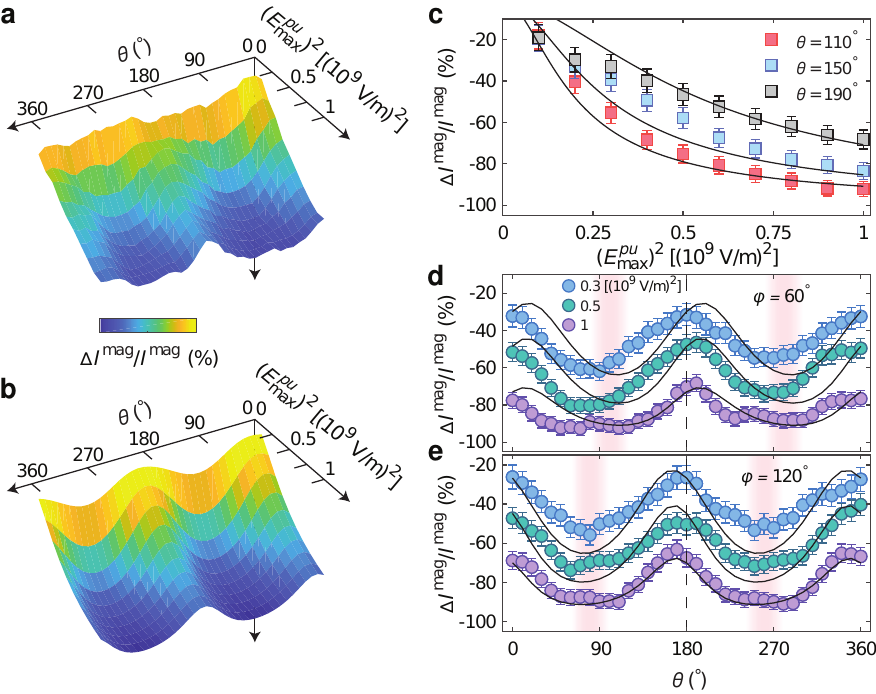}
\caption{\label{fig4} \textbf{Driving field amplitude and polarization dependence of SHG modulation.} \textbf{a,b,} Two-dimensional maps of the experimentally measured (a) and theoretically predicted (b) value of $\Delta I^\mathrm{mag}/I^\mathrm{mag}$ at time zero as a function of the peak driving field and polarization. \textbf{c,} One-dimensional cuts through the experimental map (symbols) and theoretical map (black lines) along the field axis at selected $\theta$ values. \textbf{d,e,} One-dimensional cuts along the $\theta$ axis at selected field strengths for $\varphi=60^{\circ}$ (d) and $\varphi=120^{\circ}$ (e). The positions marked by the faint vertical red bars indicate where the largest $\Delta I^\mathrm{mag}/I^\mathrm{mag}$ is predicted. The slight horizontal offset between the $\varphi=60^{\circ}$ and $\varphi=120^{\circ}$ curves arises due to a degeneracy of the final state (Supplementary section S4), not Rabi frequency anisotropy. All error bars represent the standard errors of the mean from four independent measurements.}
\end{figure*}

\clearpage
\newpage


\section*{Supplementary material}

# Supplementary Information for
# Giant modulation of optical nonlinearity by Floquet engineering

Jun-Yi Shan, M. Ye, H. Chu, Sungmin Lee, Je-Geun Park, L. Balents, and D. Hsieh

**Table of contents**

## S1. Static single ion model

The magnetic point group of MnPS$_3$ is non-centrosymmetric ($2'/m$) with symmetry generators $\langle \sigma_v, I \otimes \mathcal{T} \rangle$, or equivalently $\langle \sigma_v, C_2' \otimes \mathcal{T} \rangle$, where $\sigma_v$ is the vertical mirror perpendicular to the Mn-Mn bond, $I$ is spatial inversion, $\mathcal{T}$ is time reversal, and $C_2'$ is two-fold rotation along the Mn-Mn bond. In this symmetry group, ten independent nonzero $\chi_{ijk}^{ED(c)}$ elements are allowed as given below:

$$\chi_{ijk}^{ED(c)} = \begin{pmatrix} \begin{pmatrix} \chi_{xxx} & 0 & \chi_{xxz} \\ 0 & \chi_{xyy} & 0 \\ \chi_{xxz} & 0 & \chi_{xzz} \end{pmatrix} \\ \begin{pmatrix} 0 & \chi_{yxy} & 0 \\ \chi_{yxy} & 0 & \chi_{yyz} \\ 0 & \chi_{yyz} & 0 \end{pmatrix} \\ \begin{pmatrix} \chi_{zxx} & 0 & \chi_{zxz} \\ 0 & \chi_{zyy} & 0 \\ \chi_{zxz} & 0 & \chi_{zzz} \end{pmatrix} \end{pmatrix} \tag{S1}$$

We can compute these tensor elements quantum mechanically as

$$\chi_{ijk}^{ED(c)} \propto \sum \sum_{l,n} \left\{ \frac{\langle i|r_i|n\rangle\langle n|r_j|l\rangle\langle l|r_k|i\rangle}{(E_n - E_i - 2\hbar\omega)(E_l - E_i - \hbar\omega)} + \frac{\langle i|r_j|n\rangle\langle n|r_k|l\rangle\langle l|r_i|i\rangle}{(E_l - E_i + 2\hbar\omega)(E_n - E_i + \hbar\omega)} \right.$$
$$\left. + \frac{\langle i|r_k|n\rangle\langle n|r_i|l\rangle\langle l|r_j|i\rangle}{(E_n - E_i - \hbar\omega)(E_l - E_i + \hbar\omega)} + (j \leftrightarrow k) \right\} \tag{S2}$$

where the first sum is over Mn$^{2+}$ ions within a unit cell, $|i\rangle$ labels the system ground state, $|l\rangle$ and $|n\rangle$ are the excited states to be summed over. In the main text we determined the resonant states involved in the ED-SHG process from the optical absorption data[1], where the initial state $|i\rangle$ is a $^6A_{1g}(t_{2g}^3 e_g^2)$ state, the intermediate state $|m\rangle$ is a $^4T_{1g}(t_{2g}^4 e_g^1)$ state, and the final state $|f\rangle$ is a charge-transferred state. Taking only permutations from near resonant states, Eqn. S2 can be simplified to

$$\chi_{ijk}^{ED(c)} \propto \sum \frac{\langle i|r_i|f\rangle\langle f|r_j|m\rangle\langle m|r_k|i\rangle}{(E_f - E_i - 2\hbar\omega - i\gamma_f)(E_m - E_i - \hbar\omega)} + (j \leftrightarrow k) \quad (S3)$$

where $\gamma_f$ is a phenomenological damping rate (section S2). To compute the matrix elements in Eqn. S3, we express the zeroth-order multi-electron wavefunctions $|i_0\rangle$, $|m_0\rangle$, and $|f_0\rangle$ in the Fock space, so that the matrix elements can be calculated using the single particle orbital wavefunctions with single particle operators. Since we are using the single ion model, the basis wavefunctions are constructed according not to the crystal symmetry, but to the $Mn^{2+}$ site symmetry for a single MnPS$_3$ layer. The symmetry generators of the site magnetic point group are $C_3$ and $C_2' \otimes \mathcal{T}$, where $C_3$ is three-fold rotation along the $z$ axis. The crystal field environment around $Mn^{2+}$ is trigonal, so the original $t_{2g}$ triplet in an otherwise octahedral environment is split into an $a_1$ singlet and an $e$ doublet, where the $e$ doublet mixes with the original octahedral $e_g$ doublet, which we characterize by a parameter $\gamma$. Below we explicitly write out the five standard normalized single particle $3d$ wavefunctions in the trigonal crystal field and the three single particle $3p$ hole wavefunctions

$$\begin{aligned}
d_1^1 &= \sqrt{5}(-\frac{d_{x^2}}{2} - \frac{d_{y^2}}{2} + 2d_{z^2}) \\
d_2^\omega &= \cos\gamma\, e_1^{(1)} + \sin\gamma\, e_1^{(2)} \\
d_3^{\omega^2} &= \cos\gamma\, e_2^{(1)} + \sin\gamma\, e_2^{(2)} \\
d_4^\omega &= -\sin\gamma\, e_1^{(1)} + \cos\gamma\, e_1^{(2)} \\
d_5^{\omega^2} &= -\sin\gamma\, e_2^{(1)} + \cos\gamma\, e_2^{(2)} \\
p_1^1 &= \sqrt{3}p_z,\, p_2^\omega = \sqrt{\frac{3}{2}}(p_x - ip_y),\, p_3^{\omega^2} = \sqrt{\frac{3}{2}}(p_x + ip_y)
\end{aligned} \quad (S4)$$

where $e^{(1)} = \left\{\sqrt{\frac{15}{2}}(\frac{d_{x^2}}{2} + id_{xy} - \frac{d_{y^2}}{2}), \sqrt{\frac{15}{2}}(\frac{d_{x^2}}{2} - id_{xy} - \frac{d_{y^2}}{2})\right\}$ and $e^{(2)} = \left\{\sqrt{\frac{15}{2}}(d_{xz} - id_{yz}), \sqrt{\frac{15}{2}}(d_{xz} + id_{yz})\right\}$ are the basis wavefunctions used in constructing the mixed $d$ orbitals.

The octahedral symmetry is restored when $\cos\gamma = \sqrt{\frac{2}{3}}$ and $\sin\gamma = -\sqrt{\frac{1}{3}}$, corresponding to the absence of trigonal splitting. Here the Cartesian coordinates $\{x, y, z\}$ are aligned to be

perpendicular to the Mn-Mn bond, along the Mn-Mn bond, and along the $c$ axis. The superscript on the left-hand side of each equation corresponds to the eigenvalue under $C_3$, where $\omega$ is the complex cubic root of one. In the case of weak trigonal splitting, the states $d_1^1$, $d_2^\omega$ and $d_3^{\omega^2}$ are close in energy, so we still express them as the $t_{2g}$ manifold in Fig. 1b of the main text, while the other two $d$ orbitals are expressed as $e_g$.

Starting from zeroth-order multielectron wavefunctions constructed from the single particle basis wavefunctions defined above, we dress $|m_0\rangle$ by both the trigonal crystal field experienced by a single $Mn^{2+}$ site $H_{tri}$ and spin-orbit coupling $H_{SOC}$, and dress $|f_0\rangle$ by $H_{SOC}$, so the ED transitions from $|i\rangle$ to $|m\rangle$ and from $|m\rangle$ to $|f\rangle$ are allowed since the initial and final states must have the same spin but opposite parity[2]. $H_{tri}$ breaks local inversion symmetry and hybridizes the odd-parity $|f_0\rangle$ and the even-parity $|m_0\rangle$, whereas $H_{SOC}$ hybridizes the final state $|f_0\rangle$ which has $S = 5/2$ with a counterpart state $|f_0'\rangle$ with opposite spin on the 3$p$ hole which has $S = 3/2$. For simplicity we only consider SOC within the $p$ orbitals. We express the dressing Hamiltonians as

$$H_{tri} = \sum_{\varpi,\sigma} \eta_\sigma^\varpi p_\sigma^{\varpi\dagger} d_\sigma^\varpi + h.c. \tag{S5}$$

$$H_{SOC} = \sum_{\sigma\neq\sigma',\alpha,\beta} \lambda_{\sigma\sigma'}^{\alpha\beta} p_{\sigma,\alpha}^\dagger p_{\sigma',\beta} + h.c.$$

where $\varpi$ runs through $\{1, \omega, \omega^2\}$, $\sigma$ is the spin index and $\alpha, \beta$ label the orbitals. Following an analysis using the site magnetic point group, we show that $\eta_\sigma^1 = 0$, $\eta_\sigma^\omega = -\eta_\sigma^{\omega^2} = \eta \in \text{Im}$, and $\lambda_{\sigma\sigma'}^{\alpha\beta} \propto \lambda \in \text{Re}$. We then use the perturbation theory to derive the dressed intermediate and final states, which yields

$$|m\rangle = |m_0\rangle + \frac{H_{f_0,f_0'}^{SOC} H_{f_0',m_0}^{tri}}{(E_{m_0} - E_{f_0'})(E_{m_0} - E_{f_0})} |f_0\rangle$$

$$|f\rangle = |f_0\rangle + \frac{H_{f_0',f_0}^{SOC}}{E_{f_0} - E_{f_0'}} |f_0'\rangle \tag{S6}$$

In the single ion model the $x$ and $y$ axes are equivalent, but the stacking pattern of the bulk

MnPS$_3$ crystal induces anisotropy. We incorporate this into our model phenomenologically by explicitly forcing the matrix elements $\langle p|x|d\rangle$ and $\langle p|y|d\rangle$ to be unequal. This is accomplished by introducing anisotropy parameters $b_x \neq b_y$ and multiplying the $\langle p|x|d\rangle$ and $\langle p|y|d\rangle$ terms calculated from the isotropic model above by $b_x$ and $b_y$ respectively.

The sum in Eqn. S3 is over the two Mn$^{2+}$ ions in a unit cell. We note that the single-ion states involved in the calculation above require a finite expectation value for the $S_z$ quantum number, either up or down. Therefore, if the spins are disordered no ED-SHG is allowed. If the spins are ordered, then $\chi_{ijk}^{ED(c)}$ is proportional to the AFM order parameter, i.e., $\langle S_{z,1}\rangle - \langle S_{z,2}\rangle$.

Since the site symmetry we used is higher than the crystal symmetry, our single ion model only provides a subset of the tensor element values in Eqn. S1, which we list below.

$$\begin{aligned}
\chi_{xxx}^{ED(c)} &= i(\langle S_{z,1}\rangle - \langle S_{z,2}\rangle)\lambda^2 \eta^{(i)} b_x^3 \cos\gamma\,(\sqrt{6}\cos\gamma - 6\cos 2\gamma)/15\sqrt{10} \\
\chi_{xxz}^{ED(c)} &= i(\langle S_{z,1}\rangle - \langle S_{z,2}\rangle)\lambda^2 \eta^{(i)} b_x^2 \sin 2\gamma\,(3\sqrt{2}\cos\gamma - 2\sqrt{3})/30\sqrt{5} \\
\chi_{xyy}^{ED(c)} &= i(\langle S_{z,1}\rangle - \langle S_{z,2}\rangle)\lambda^2 \eta^{(i)} b_x b_y^2 \cos\gamma\,(6\cos 2\gamma - \sqrt{6}\cos\gamma)/15\sqrt{10} \\
\chi_{yxy}^{ED(c)} &= i(\langle S_{z,1}\rangle - \langle S_{z,2}\rangle)\lambda^2 \eta^{(i)} b_x b_y^2 \cos\gamma\,(6\cos 2\gamma - \sqrt{6}\cos\gamma)/15\sqrt{10} \\
\chi_{yyz}^{ED(c)} &= i(\langle S_{z,1}\rangle - \langle S_{z,2}\rangle)\lambda^2 \eta^{(i)} b_y^2 \sin 2\gamma\,(3\sqrt{2}\cos\gamma - 2\sqrt{3})/30\sqrt{5}
\end{aligned} \quad (S7)$$

where the energy denominators have been dropped for clarity. However, this subset suffices to explain our static RA-SHG data since the ED-SHG amplitude in P$_{in}$-P$_{out}$ geometry only contains these elements as shown below

$$\begin{aligned}
P^{(c)}(2\omega) = \cos^3\theta_i \left[\chi_{xxx}^{ED(c)}\cos^3\varphi + \left(2\chi_{yxy}^{ED(c)} + \chi_{xyy}^{ED(c)}\right)\cos\varphi\sin^2\varphi\right] \\
- 2\cos^2\theta_i \sin\theta_i\,(\chi_{xxz}^{ED(c)}\cos^2\varphi + \chi_{yyz}^{ED(c)}\sin^2\varphi)
\end{aligned} \quad (S8)$$

where $\theta_i$ is the angle of incidence of the SHG probe beam and $\varphi$ is the azimuthal angle.

## S2. Determination of $E_g$ and $\gamma_f$

To extract the optical gap $E_g$ of MnPS$_3$, we fit our measured optical absorption data to a general model for the interband absorption coefficient K due to direct allowed transitions in three-dimensional semiconductors[3] $K(E) = (A/E)\sqrt{E - E_g}\,\Theta(E - E_g)$. Here $\Theta$ is the Heaviside function, $E$ is the photon energy and $A$ is a constant. A best fit (Fig. S1) yields $E_g$ = 3.09 ± 0.04 eV. This is in good agreement with a DFT calculation using HSE06 functionals[4], which yields $E_g$ = 3.08 eV. This justifies our treating the SHG photon energy $2\hbar\omega$ = 3.1 eV as being resonant with the unperturbed $E_g$.

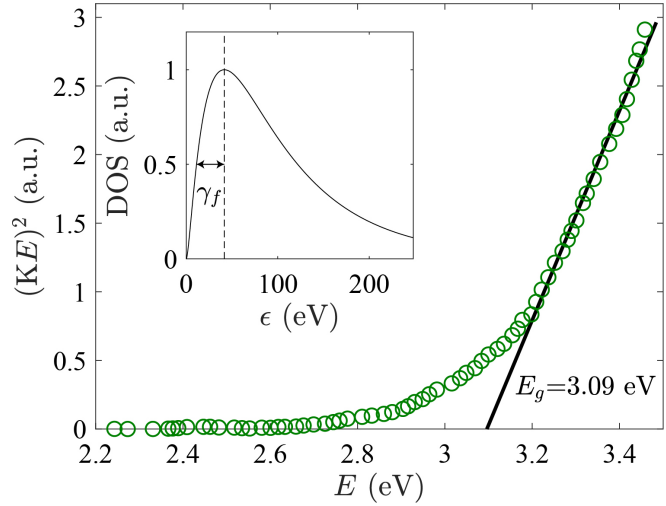

Fig. S1. Optical absorption data. The relationship between $(KE)^2$ and $E$ is plotted (green circles) to facilitate the linear fit (black curve). Inset shows the DOS of in-gap impurity states.

To estimate the phenomenological damping rate $\gamma_f$ used in Eqn. 1 of the main text, we assume that the broadening of the absorption edge is dominated by in-gap impurity states with a density of states DOS proportional to $(\epsilon/\epsilon_0)^2 \exp(-\sqrt{\epsilon/\epsilon_0})$ (inset Fig. S1) following Ref. [3], where $\epsilon_0$ is the characteristic energy scale corresponding to the ionization energy of the impurities. Here $\epsilon_0$ was estimated[3] to be 21 cm$^{-1}$, or 2.6 meV for MnPS$_3$, which is a typical value for semiconductors[5]. We approximate $\gamma_f$ from the half width half maximum of the DOS curve as shown in the inset of Fig. S1, which is around 30 meV.

## S3. Linear coupling of $\chi_{ijk}^{ED(c)}$ to the AFM order parameter

To confirm the validity of our static single-ion SHG model, we checked whether $\chi_{ijk}^{ED(c)}$ linearly couples to the AFM order parameter. We extract $\chi_{ijk}^{ED(c)}$ from the temperature dependence of the static SHG intensity $I(2\omega)$ (main text Fig. 1d) by assuming that $I(2\omega)$ is the square of the coherent sum of an $i$-type induced polarization $P^{(i)}$, which is constant over the whole temperature range, and a $c$-type induced polarization $P^{(c)}$, which is proportional to $\chi_{ijk}^{ED(c)}$. We assume a critical scaling behavior of the form $\chi_{ijk}^{ED(c)} \propto (T_N - T)^\beta$. Fits to this form show that in the vicinity of $T_N$ (72 K $< T <$ 78 K), the fitted $\beta$ is $0.32 \pm 0.02$, while at lower temperatures (30 K $< T <$ 72 K), the fitted $\beta$ is $0.24 \pm 0.03$ (Fig. S2). This is consistent with neutron scattering measurements[6] which observed a crossover from $\beta = 0.32$ to $\beta = 0.25$, confirming that $\chi_{ijk}^{ED(c)}$ linearly couples to the AFM order parameter.

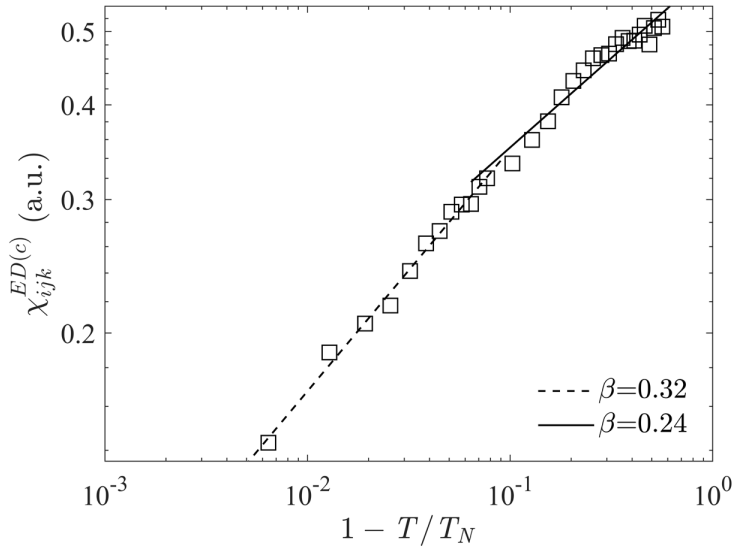

Fig. S2. The log-log plot of the critical behavior of $\chi_{ijk}^{ED(c)}$ (squares). Linear fits within two different temperature ranges are overlaid (lines).

## S4. Dynamical Floquet model

In treating the time-dependent problem we neglect the pulse envelope, since the characteristic time scale of the system $h/E_g \approx 1$ fs is much shorter than our typical Gaussian pulse widths. We also note that using our experimental values of $\hbar\Omega = 0.66$ eV and $E_{max}^{pu} = 10^9$ V/m, the effective Rabi frequency $\hbar\Omega^R$ is 0.4 eV and the ponderomotive energy under the effective-mass approximation[7] is 0.09 eV, which is much smaller than both $\hbar\Omega$ and $\hbar\Omega^R$. Therefore, we neglect the term quadratic in $E^{pu}$. Moreover, we confine ourselves to the subspace spanned by $|i\rangle$ and $|f\rangle$ because 1) the coupling between them is electric dipole allowed, 2) these two states are resonant with our SHG photon energy and their modulations are readily observed as a modulation in SHG. Coupling of $|m\rangle$ to higher energy states is possible, but the effects are negligible since their energy difference is more off resonant from the drive. Also, we find that even if the $|m\rangle$ state is hypothetically shifted down by an amount $\Delta E_{max} \approx 100$ meV as is the case for $|i\rangle$, it would only slightly change the maximum $\Delta I^{mag}/I^{mag}$ from −91 % to −88 %.

We note that in order for our analysis using $|i\rangle$ and $|f\rangle$ as pure states to be valid (main text Eqn. 3), the lifetime of these states must be sufficiently long. Otherwise, one must use density matrices to express the states, and our analysis would be invalid. It has been shown that[8] in the limit that $\gamma_f \ll \Omega^R$, where $\Omega^R$ is the Rabi frequency, the off-diagonal elements in the density matrices can be neglected and analysis using pure states is valid. In our case, $\gamma_f$ is around 30 meV and $\Omega^R$ is an order of magnitude larger, so the Floquet picture is valid. We do not consider the intermediate state lifetime because it is not a relevant state in this Floquet problem.

Starting from the time-dependent Hamiltonian $H(t) = H_0 + e\vec{r} \cdot \vec{E}^{pu} \cos\Omega t$, the infinite dimensional Floquet Hamiltonian $(H_F)_{mn} = \frac{1}{2\pi/\Omega} \int_0^{2\pi/\Omega} e^{i\Omega t(m-n)} H(t) - n\hbar\Omega\delta_{mn}$ can be expressed in matrix form as

$$H_F = \begin{pmatrix} \ddots & & & & \\ & H_0 - \hbar\Omega & H_1 & & \\ & H_1 & H_0 & H_1 & \\ & & H_1 & H_0 + \hbar\Omega & \\ & & & & \ddots \end{pmatrix} \quad (S9)$$

where $H_0$ is the static Hamiltonian $H_0 = \begin{pmatrix} E_i & 0 \\ 0 & E_f \end{pmatrix}$ and $H_1$ is the coupling term $H_1 = \begin{pmatrix} 0 & \hbar\Omega^R \\ \hbar\Omega^R & 0 \end{pmatrix}$. The Rabi frequency $\Omega^R$ is proportional to $E^{pu}$. Due to the phenomenological in-plane anisotropy introduced in section S1, there is also an in-plane anisotropy of $\Omega^R$. In the dynamical model we use the same ratio $b_y/b_x \sim 1.5$ as fitted from the static RA patterns.

The time-dependent eigenstates of $H(t)$ can be constructed from the eigenvalues and eigenvectors of $H_F$, using a general solution of the form[9],

$$|\psi_n(t)\rangle = e^{-i\varepsilon_n t/\hbar} \sum_m e^{-im\Omega t} |\phi_n^{(m)}\rangle \quad (S10)$$

$$= e^{-i\varepsilon_n t/\hbar} \sum_m e^{-im\Omega t} (c_{n,i}^{(m)}|i + m\hbar\Omega\rangle + c_{n,f}^{(m)}|f + m\hbar\Omega\rangle)$$

where $\varepsilon_n$ is the $n^{\text{th}}$ eigenvalue of $H_F$ and $(\ldots, c_{n,i}^{(-1)}, c_{n,f}^{(-1)}, c_{n,i}^{(0)}, c_{n,f}^{(0)}, c_{n,i}^{(1)}, c_{n,f}^{(1)}, \ldots)$ is the corresponding eigenvector, where $m$ corresponds to the order of the Floquet sector. There are two components for each Floquet sector because $H_0$ is two dimensional, corresponding to the unperturbed $|i\rangle$ and $|f\rangle$.

To construct $|i'\rangle$ given in Eqn. 3 of the main text, we first determine $E_{i'} = E_i - \Delta E$ from the eigenvalues of $H_F$. The corresponding normalized eigenvector to the eigenvalue $E_{i'}$ is dominated by the component $c_{n,i}^{(0)}$, which corresponds to the unperturbed $|i\rangle$ in the zeroth Floquet sector, and we define $c_{n,i}^{(0)}$ to be $\cos\alpha$. Other $c_{n,i}^{(2k)}$ can be nonzero, but they are negligible compared to $\cos\alpha$. All $c_{n,i}^{(2k+1)}$ components are zero. Instead, all $c_{n,f}^{(2k+1)}$ components are nonzero, and these components all correspond to the same unperturbed $|f\rangle$ electronic state. Therefore, we sum up these contributions, each of them is a real coefficient

$c_{n,f}^{(2k+1)}$ times its corresponding phase $e^{-im\Omega t}$, and we define the summation as $\sin\alpha\, e^{i\phi(t)}$. Perturbatively, $\sin\alpha$ is proportional to $E^{pu}$ (Fig. S3). Lastly, all $c_{n,f}^{(2k)}$ components of the eigenvector are zero. Therefore, we have

$$|i'\rangle = e^{-i(E_i-\Delta E)t/\hbar}(\cos\alpha\,|i\rangle + \sum_k e^{-i(2k+1)\Omega t}\,c_f^{(2k+1)}|f\rangle) \qquad (S11)$$
$$= e^{-i(E_i-\Delta E)t/\hbar}(\cos\alpha\,|i\rangle + \sin\alpha\, e^{i\phi(t)}|f\rangle)$$

While the $|i\rangle$ component has an energy of $E_i - \Delta E$ and is still nearly resonant in the SHG process, the mixed-in $|f\rangle$ components have energies of $E_i - \Delta E + (2k+1)\hbar\Omega$, which are far from resonance in the SHG process. Therefore, in the dynamical SHG calculations we only keep the $\cos\alpha\,|i\rangle$ term and neglect the $\sin\alpha\, e^{i\phi(t)}|f\rangle$ term. The same holds true for the components in the driven $|f'\rangle$ state.

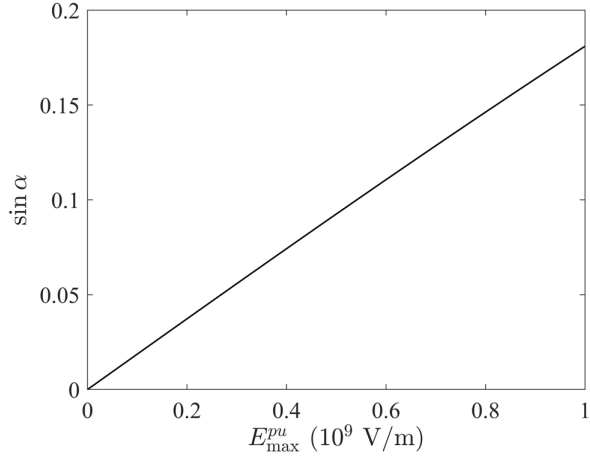

Fig. S3. The mixing factor $\sin\alpha$ as a function of $E_{\max}^{pu}$.

With the creation of Floquet sidebands as shown in Fig. 2a of the main text, an SHG spectrum would also show resonances due to these sidebands (Fig. S4). The time-averaged wavefunction amplitude of the $n$-th order sideband is approximately equal to $[\hbar\Omega^R/(E_f - E_i)]^{|n|}$. The lowest-order resonance involving the sidebands occurs at $2\hbar\omega = E_{f'} - E_{i'} \pm 2\hbar\Omega$, because the transitions involving energy $E_{f'} - E_{i'} \pm \hbar\Omega$ (e.g., from the photon-dressed states $|i' \pm \hbar\Omega\rangle$ to $|f'\rangle$) are forbidden due to parity selection rules. Since the sideband wavefunction occurs in $\chi_{ijk}^{ED(c)}$ twice (Eqn. S3), the magnitude of $\chi_{ijk}^{ED(c)}$ at the $2\hbar\omega = E_{f'} - E_{i'} \pm 2\hbar\Omega$ resonances is weaker than the main resonance at $2\hbar\omega = E_{f'} - E_{i'}$ by a factor of

$[\hbar\Omega^R/(E_f - E_i)]^4$. For $E_{max}^{pu} = 10^9$ V/m, the resonances due to the Floquet sidebands would be weaker than the main resonance by $10^{-4}$.

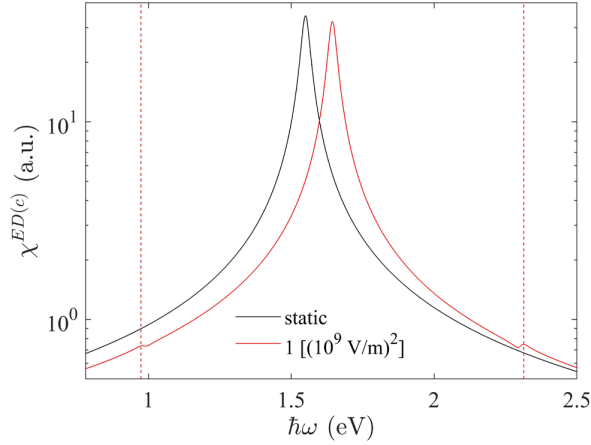

Fig. S4. The theoretical SHG spectrum within the single ion picture for both the static case and the $(E_{max}^{pu})^2 = 1\ [(10^9\ \text{V/m})^2]$ case. Resonances due to Floquet sidebands are indicated by the red dashed line. Resonances involving the $|m\rangle$ state are ignored.

For the driving frequencies in Fig. 3b of the main text, their respective $\Delta E_{max}$ and $\sin\alpha$ with $E_{max}^{pu} = 10^9$ V/m drive are shown in Table S1.

| $\hbar\Omega$ (eV) | $\Delta E_{max}$ (meV) | peak $\sin\alpha$ (%) |
|---|---|---|
| 0.66 | 94 | 18.1 |
| 0.99 | 100 | 19.5 |
| 1.30 | 107 | 21.1 |
| 1.55 | 116 | 23.2 |

Table S1. Floquet engineering with different driving frequencies

Because these frequencies lead to similar $\Delta E_{max}$ and $\sin\alpha$, they result in similar modulation to $I^{mag}$, consistent with results reported in Fig. 3b of the main text. In theory $I^{mag}$ can be fully suppressed with larger driving amplitudes than what we used, but practically the material becomes damaged, possibly through multi-photon ionization (Keldysh parameter ~ 3 at $E_{max}^{pu} = 10^9$ V/m).

Figures 4d and 4e of the main text show that the SHG intensity varies as a function of $\theta$ and peaks around 0° and 180°. However, there exists a small but finite offset from 0° and 180°, which is different for $\varphi = 60°$ and 120°. The reason is as follows. The anisotropy in the polarization angle $\theta$ comes from two sources. The major source is the Rabi frequency anisotropy due to the layer stacking, which produces peaks at exactly 0° and 180°. The secondary source comes from the fact that the $|f\rangle$ state is not a single state but is actually a degenerate manifold of states - the excited sulfur hole and manganese electron can reside in different $3p$ and $t_{2g}$ orbitals respectively (main text Fig. 1b). Upon driving, a different energy shift and mixing factor is induced between the $|i\rangle$ state and each of the different $|f\rangle$ states. Since these values all depend on the orientation of the driving field, this effect produces an additional modulation of the SHG intensity with $\theta$. Combined with the previous source, this produces intensity peaks shifted slightly away from 0° and 180°. The coupling of the SHG probe field to each of the different $|f\rangle$ states changes depending on the polarization of the SHG probe beam $\varphi$. As a result, the $\theta$ dependence of the SHG intensity induced by the degeneracy of the $|f\rangle$ states will change depending on $\varphi$, which explains why the curves in Fig. 4d are slightly different from the curves in Fig. 4e.

## S5. Transient fundamental and sum-frequency generation response

The Bloembergen-Pershan correction[10] relates the measured SHG susceptibility in reflection geometry $\chi^R$ to the intrinsic SHG susceptibility $\chi$ via $\frac{\chi^R}{\chi} = [\frac{2}{n(\omega)+1}]^2 \frac{1}{[n(2\omega)+n(\omega)][n(2\omega)+1]}$, where $n$ is the index of refraction. In order to rule out the trivial possibility that the $I^{mag}$ drop is caused by changes to $n$ in the Bloembergen-Pershan correction, we measured changes to the linear reflectivity $\Delta R/R$ at both the fundamental (1.55 eV, Fig. S5) and second harmonic (3.1 eV) frequencies caused by the 0.66 eV drive. With 3.1 eV probe, $\Delta R/R$ is always below our noise level and so $\Delta n(2\omega)/n(2\omega) \approx 0$. With 1.55 eV probe, the largest measured $\Delta R/R$ is -2.5 %, which corresponds to $\Delta n(\omega)/n(\omega)$ of -1.2 % using the Fresnel equations. These lead to a change in the Bloembergen-Pershan correction of +2.1 %. This is not only far too small to explain the over 90 % drop of the SHG intensity but is also of opposite sign. Therefore, the observed SHG modulation must predominantly originate from changes to $\chi$.

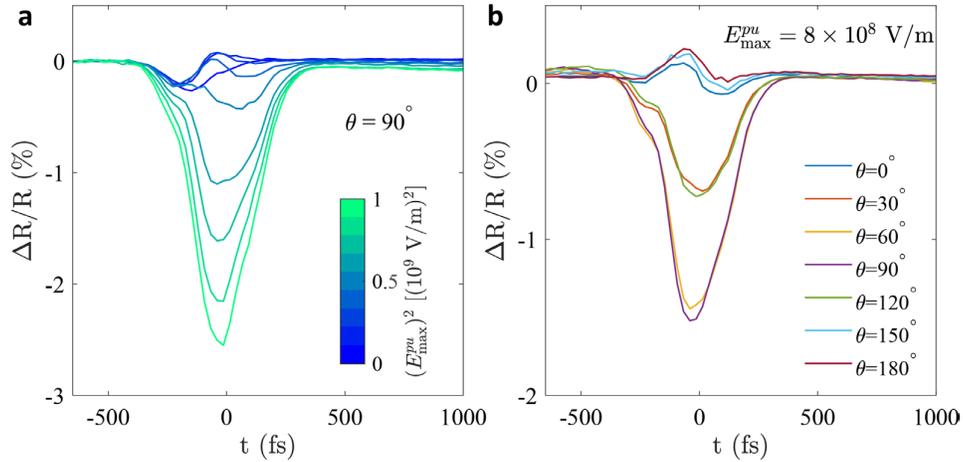

Fig. S5. $\Delta R/R$ with 1.55 eV probe taken at various driving **a**, amplitudes and **b**, polarizations.

We note that the driving amplitude and polarization dependences of $\Delta R/R$ at 1.55 eV are similar to those of $\Delta I^{mag}/I^{mag}$, so the observed $\Delta R/R$ is likely also a result of Floquet engineering. To show this, we mathematically calculated the Floquet driving induced $\Delta R/R$. The quantum mechanical expression for the linear electric susceptibility is:

$$\chi_{ij} \sim \frac{\langle i|r_i|m\rangle\langle m|r_j|i\rangle}{E_m - E_i - \hbar\omega} \tag{S12}$$

Under Floquet driving conditions, $E_i$ shifts down by $\Delta E$ and the state $|i\rangle$ gains a mixing factor $\cos\alpha$. The refractive index $n$ can be computed from $\chi$ using the relation $n^2 = \varepsilon = 1 + \chi$, where $\varepsilon$ is the relative permittivity. The linear reflectivity $R$ can then be obtained from the refractive index $n$ through the Fresnel formula. For small driving amplitudes, we show that to lowest order:

$$\frac{\Delta R}{R} \propto -\left(\frac{1}{2}\alpha^2 + \frac{\Delta E}{E_m - E_i - \hbar\omega}\right) \tag{S13}$$

Since $\alpha \propto E^{pu}$ and $\Delta E \propto (E^{pu})^2$, both the state mixing and level shift contributions to the differential reflectivity are quadratic in $E^{pu}$.

When the pump and probe pulses overlap, there is sum-frequency generation (SFG) $P_i(\omega + \Omega) = \chi_{ijk}^{SF} E_j^{pr}(\omega) E_k^{pu}(\Omega)$, with $\hbar\omega = 1.55$ eV, $\hbar\Omega = 0.66$ eV, and $\hbar\omega + \hbar\Omega = 2.21$ eV in our experimental setting. To rule out the possibility that the SFG process competes with SHG[11,12] and causes the SHG drop, we measured the SFG intensity as a function of the driving field amplitude (Fig. S6).

For all probe fluences used, we observe that upon increasing the driving amplitude the SFG intensity exhibits a non-monotonic dependence on $E_{max}^{pu}$, first increasing with $E_{max}^{pu}$ and then decreasing. This rules out the possibility that $\chi_{ijk}^{SF}$ is enhanced by the drive. On the contrary, the $I^{SF}$ trend can be well explained by an inverse power law-like suppression of $\chi_{ijk}^{SF}$ as a function of $E_{max}^{pu}$, analogous to the case for $\chi_{ijk}^{ED(c)}$, showing a consistent modulation of second-order optical nonlinearities by the drive field.

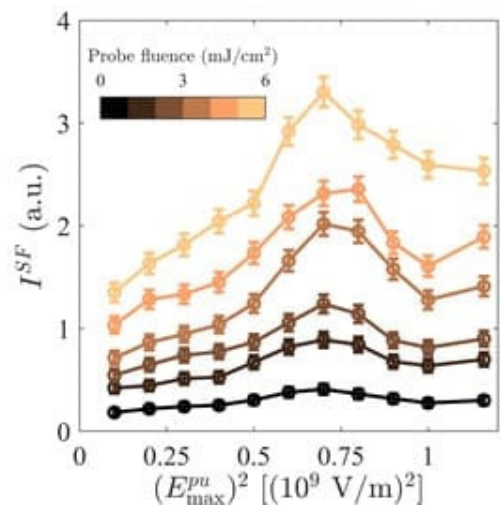

Fig. S6. Driving field amplitude dependence of SFG intensity at various probe fluences.

## S6. Time-resolved SHG data at 70 K and 90 K

To place an upper bound on the heating due to the sub-gap $\hbar\Omega = 0.66$ eV drive, we performed time-resolved SHG measurements at 70 K where the SHG intensity shows strong dependence on temperature (Fig. 1d). As shown in Fig. S7a the pre- and post- time zero SHG values are equal within our noise, which allows us to bound the transient temperature rise to less than 1 K.

To confirm that the SHG intensity modulation with $\hbar\Omega = 0.66$ eV driving is due to a modulation of only the $\chi_{ijk}^{ED(c)}$ tensor, we measured the effect of driving on the SHG intensity above $T_N$ at 90 K (Fig. S7b) that originates exclusively from higher multipole processes. We observed no change upon driving. Therefore, we attribute all changes in the SHG intensity to $\Delta I^{mag}$.

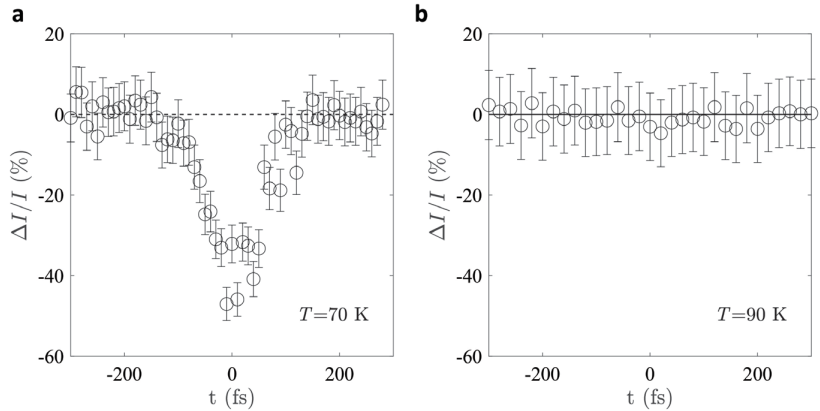

Fig. S7. Time-resolved SHG measured at **a,** 70 K and **b,** 90 K with $\hbar\Omega = 0.66$ eV driving. $E_{max}^{pu} = 10^9$ V/m, $\theta = 90°$ and $\varphi = 60°$.

## S7. Proof of the heating picture for $\hbar\Omega = 2.07$ eV drive

The heating scenario is verified in two ways. First, the time-resolved RA patterns taken at 10 K within the exponential decay time window can be directly mapped onto static RA patterns taken at higher temperatures (Fig. S8).

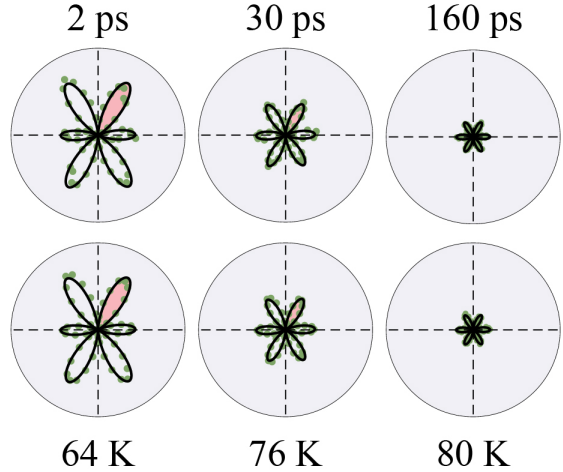

Fig. S8. Upper panels show the time-resolved RA patterns taken at 10 K, and lower panels show the static RA patterns taken at higher temperatures. Identical fits to the crystal point group for each column are overlaid (black lines).

Second, within the heating picture, we can estimate $I^{mag}$ at long time delays by assuming an equilibrated temperature within the probed region and mapping the absorption induced temperature rise $\Delta T$ to the static $I^{mag}$ curve (main text Fig. 1d). Using the reported MnPS$_3$ specific heat[13] $c$ and absorption coefficient at 600 nm, which corresponds to a penetration depth of around 100 μm[14], we can estimate $\Delta T$ for each driving amplitude calculated from pulse energies $Q$ by $Q = \int_{T_0}^{T_0+\Delta T} c(T)dT$, where $T_0$ is our base temperature. We show that the calculated $E_{max}^{pu}$ dependence of $\Delta I^{mag}/I^{mag}$ qualitatively agrees with experiment (Fig. S9). The slope of the $\Delta I^{mag}/I^{mag}$ curve at low $E_{max}^{pu}$ is small because the static $I^{mag}$ is almost constant at low temperatures. The slope of the $\Delta I^{mag}/I^{mag}$ curve at higher $E_{max}^{pu}$ is larger, in accordance with the highly temperature dependent $I^{mag}$ near $T_N$. When the effective temperature $T_0 + \Delta T$ exceeds $T_N$, $I^{mag}$ becomes zero. We note that the calculated curve in Fig. S9 is based on an extremely simplistic model that assumes a uniformly heated cylinder of material, which neglects spatial gradients of the pump and probe beams, heat

diffusion and thermally induced structural changes that may affect the EQ SHG contribution. These are likely to be the main reasons for why the theoretical curve and the experimental data do not quantitatively match.

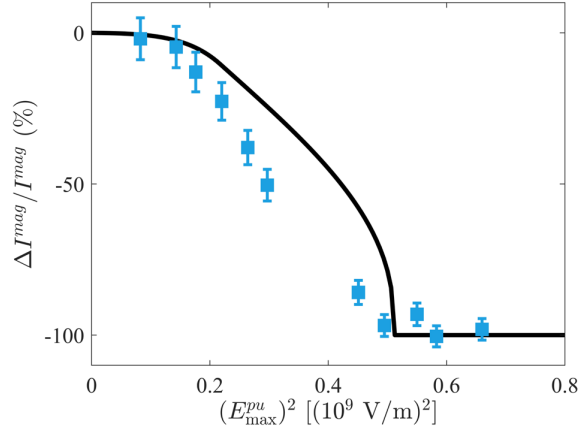

Fig. S9. The calculated (black) and experimental (blue) driving amplitude dependence of $\Delta I^{mag}/I^{mag}$ plateau values. Experimental data are taken at $t = 200$ ps.

Finally, to estimate the time for heat diffusion out of the excited area $t_d$, we calculated the thermal diffusivity $D_{th} = k/(\rho c)$, where $k$ is the thermal conductivity, $\rho$ is the density and $c$ is the specific heat[13,15,16]. With $k = 6.3$ W/m K, $\rho = 2680$ kg/m³, $c = 330$ J/ kg K, yielding $D_{th} = 7.1 \times 10^{-6}$ m²/s, we estimated that $t_d$ is on the order of μs. This explains the $\Delta I^{mag}/I^{mag}$ plateau at long time delays.

## S8. Calculation of photon-assisted hopping contribution

Our single ion model assumes well-localized states and neglects the band curvature. To estimate effects of band structure renormalization, we calculated the effects of photon-assisted hopping and dynamical localization within a Hubbard model[17]. In this picture the Floquet drive renormalizes the hopping parameter $t$ to[18]

$$t_{eff} = t \sqrt{\sum_{n=-\infty}^{+\infty} \frac{J_{|n|}(\mathcal{E})^2}{1 + n\Omega/U}} \tag{S14}$$

where $J_{|n|}$ is the $|n|$-th order Bessel function, $\mathcal{E}$ is the Floquet parameter and $U$ is the on-site Coulomb interaction parameter. Using a peak driving value of $\mathcal{E} = 0.5$, $\Omega = 0.66$ eV and $U = 5$ eV[19], we estimate that $t_{eff}/t = 1.0013$. Assuming a valence bandwidth of 300 meV[4], the $t$ amplitude modulation roughly corresponds to an energy scale of 0.4 meV, which is much smaller than the ~ 100 meV level shifts caused by the single ion effects.

## S9. Energy-dispersive x-ray spectroscopy and magnetization data

We performed energy-dispersive x-ray spectroscopy (EDX, Fig. S10) to confirm the sample stoichiometry. We also measured the temperature dependence of the sample magnetic susceptibility using a SQUID magnetometer (Fig. S11), which shows excellent agreement with previously published results[20].

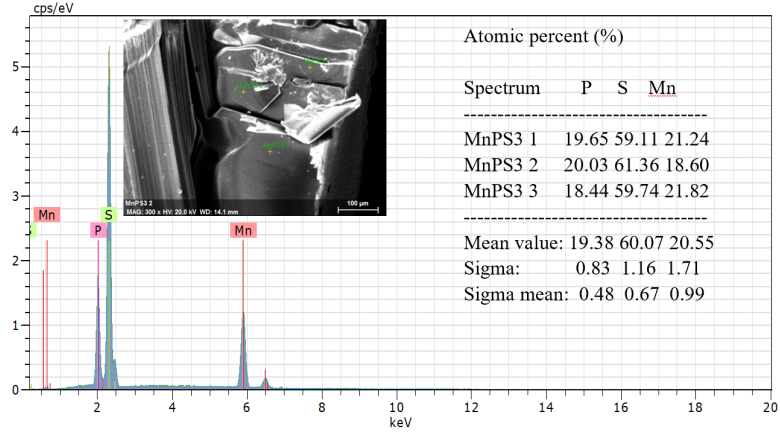

Fig. S10. The EDX spectrum and the calculated atomic percentage measured at three different spots.

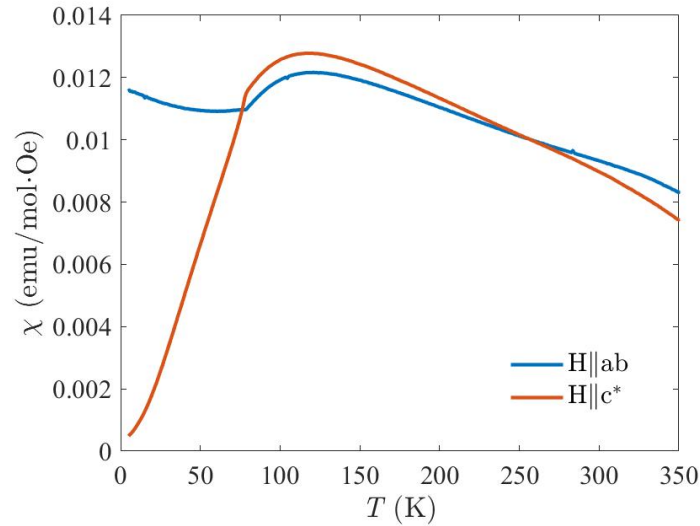

Fig. S11. The magnetic susceptibility measured with the magnetic field parallel to the *ab* plane and to the *c\** axis, which is the out-of-plane direction.





\end{document}